\documentstyle[11pt,appb,epsf,epsfig]{article} 
\def\beq{\begin{equation}}
\def\eeq{\end{equation}}
\def\beqar{\begin{eqnarray}}
\def\eeqar{\end{eqnarray}}
\def\barr#1{\begin{array}{#1}}
\def\earr{\end{array}}
\def\bfi{\begin{figure}}
\def\efi{\end{figure}}
\def\btab{\begin{table}}
\def\etab{\end{table}}
\def\bce{\begin{center}}
\def\ece{\end{center}}

\def\text{\textstyle}

\def\al{\alpha}

\def\de{\delta}

\def\De{\Delta}

\def\refeq#1{\mbox{eq.~(\ref{#1})}}
\def\reffi#1{\mbox{Fig.~\ref{#1}}}

\def\refta#1{\mbox{Tab.~\ref{#1}}}
\def\citere#1{\mbox{Ref.~\cite{#1}}}

\newcommand{\GeV}{\unskip\,{\mathrm GeV}}

\newcommand{\TeV}{\unskip\,{\mathrm TeV}}

\def\mathswitchr#1{\relax\ifmmode{\mathrm{#1}}\else$\mathrm{#1}$\fi}

\newcommand{\PW}{\mathswitchr W}
\newcommand{\PZ}{\mathswitchr Z}

\newcommand{\PH}{\mathswitchr H}

\newcommand{\Pb}{\mathswitchr b}

\newcommand{\Pt}{\mathswitchr t}

\def\mathswitch#1{\relax\ifmmode#1\else$#1$\fi}

\newcommand{\MW}{\mathswitch {M_\PW}}

\newcommand{\MZ}{\mathswitch {M_\PZ}}
\newcommand{\MH}{\mathswitch {M_\PH}}

\newcommand{\Mb}{\mathswitch {m_\Pb}}
\newcommand{\Mt}{\mathswitch {m_\Pt}}

\newcommand{\scrs}{\scriptscriptstyle}
\newcommand{\sw}{\mathswitch {s_{\scrs\PW}}}

\newcommand{\cw}{\mathswitch {c_{\scrs\PW}}}

\newcommand{\GF}{\mathswitch {G_\mu}}

\def\tb{\tan\beta}
\newcommand{\CTb}{\cot \beta}

\newcommand{\se}{self-en\-er\-gy}
\newcommand{\ses}{self-en\-er\-gies}
\newcommand{\fea}{{\em FeynArts}}

\newcommand{\two}{{\em TwoCalc}}

\renewcommand{\Re}{\mathop{\mathrm{Re}}}

\def\draftdate{\relax}
\def\mda{\relax}
\def\mua{\relax}
\def\mla{\relax}
\def\draft{
\def\thtystars{******************************}
\def\sixtystars{\thtystars\thtystars}
\typeout{}
\typeout{\sixtystars**}
\typeout{* Draft mode!
         For final version remove \protect\draft\space in source file
*}
\typeout{\sixtystars**}
\typeout{}
\def\draftdate{\today}
\def\mua{\marginpar[\boldmath\hfil$\uparrow$]%
                   {\boldmath$\uparrow$\hfil}%
                    \typeout{marginpar: $\uparrow$}\ignorespaces}
\def\mda{\marginpar[\boldmath\hfil$\downarrow$]%
                   {\boldmath$\downarrow$\hfil}%
                    \typeout{marginpar: $\downarrow$}\ignorespaces}
\def\mla{\marginpar[\boldmath\hfil$\rightarrow$]%
                   {\boldmath$\leftarrow $\hfil}%
                    \typeout{marginpar:
$\leftrightarrow$}\ignorespaces}
\def\Mua{\marginpar[\boldmath\hfil$\Uparrow$]%
                   {\boldmath$\Uparrow$\hfil}%
                    \typeout{marginpar: $\Uparrow$}\ignorespaces}
\def\Mda{\marginpar[\boldmath\hfil$\Downarrow$]%
                   {\boldmath$\Downarrow$\hfil}%
                    \typeout{marginpar: $\Downarrow$}\ignorespaces}
\def\Mla{\marginpar[\boldmath\hfil$\Rightarrow$]%
                   {\boldmath$\Leftarrow $\hfil}%
                    \typeout{marginpar:
$\Leftrightarrow$}\ignorespaces}
\overfullrule 5pt
\oddsidemargin -15mm
\marginparwidth 29mm
}

\begin{document}
\null
\hfill KA-TP-19-1997\\
\null
\hfill hep-ph/9711254\\
\vskip .8cm
\begin{center}
{\Large \bf Higher-order corrections\\[.5em]
to precision observables\\[.5em]
in the Standard Model and the MSSM%
\footnote{Talk presented at the XXI School of Theoretical Physics, 
``Recent progress in theory and phenomenology of fundamental interactions'', 
Ustron, September 19--24, 1997.}
}
\vskip 2.5em
{\large
{\sc Georg Weiglein}\\[1ex]
{\normalsize \it Institut f\"ur Theoretische Physik, Universit\"at
Karlsruhe,
D-76128 Karlsruhe, Germany}
}
\vskip 2em
\end{center} \par
\vskip 1.2cm
\vfil
{\bf Abstract} \par
A summary of recent results obtained for higher-order corrections to
precision observables in the Standard Model and the Minimal
Supersymmetric Standard Model is given. In the Standard Model
electroweak two-loop results valid for arbitrary values of the masses
of the top quark, the Higgs boson and the gauge bosons are discussed.
For the example of two specific diagrams the exact two-loop result is
compared with the result of an expansion in the top-quark mass up to
next-to-leading order. Furthermore the Higgs-mass dependence of the
two-loop corrections to the relation between the gauge-boson masses
is analyzed. In the MSSM the exact gluonic corrections to $\Delta r$
are derived. They are compared with the leading contribution entering
via the $\rho$~parameter.
\par
\vskip 1cm
\null
\setcounter{page}{0}
\clearpage

%% \eqsec  % uncomment this line to get equations numbered by (sec.num)
%\title{HIGHER-ORDER CORRECTIONS TO\\
%PRECISION OBSERVABLES IN\\
%THE STANDARD MODEL AND THE MSSM\thanks{Presented at the XXI School of
%Theoretical Physics, Recent progress in theory and phenomenology of
%fundamental interactions, Ustron, September 19--24, 1997.}
%% you can use \\ to break lines
%}
%\author{Georg Weiglein
%\address{Institut f\"ur Theoretische Physik, Universit\"at Karlsruhe,
%D--76128 Karlsruhe, Germany}
%}
%\maketitle
%\begin{abstract}
%A summary of recent results obtained for higher-order corrections to
%precision observables in the Standard Model and the Minimal
%Supersymmetric Standard Model is given. In the Standard Model 
%electroweak two-loop results valid for arbitrary values
%of the masses of the top quark, the Higgs boson and the gauge bosons
%are discussed.
%For the example of two specific diagrams the exact two-loop
%result is compared with the result of an expansion in the top-quark
%mass up to next-to-leading order. Furthermore the Higgs-mass
%dependence of the two-loop corrections to the relation between the
%gauge-boson masses is analyzed. In the MSSM the 
%gluonic corrections to $\Delta r$ are derived. They are compared with the 
%leading contribution entering via the $\rho$~parameter.
%
%\end{abstract}
%\PACS{12.15.Ji, 12.15.Lk, 12.60.Jv, 13.35.Bv}
  
\section{Introduction}

The experimental accuracy meanwhile reached for the electroweak precision 
observables allows to test the theory, namely the electroweak Standard 
Model (SM) and
its extensions, most prominently the Minimal Supersymmetric Standard
Model (MSSM), at the quantum level, where all parameters of the model
enter the theoretical predictions. 

In this way one is able to set constraints in the SM on the mass of the 
Higgs boson, which is the last missing ingredient of the minimal SM. 
{}From the most recent 
global SM fits to all available data one obtains an upper bound
for the Higgs-boson mass of $420$~GeV at $95\%$ C.L.~\cite{datasum97}.
This bound is considerably affected by the error in the theoretical 
predictions due to missing higher-order corrections, which gives rise to an
uncertainty of the upper bound of about $100$~GeV.
The main uncertainty in this context comes from the electroweak
two-loop corrections, for which the results obtained so far have
been restricted to expansions for asymptotically large values of the 
top-quark mass, $\Mt$, or the Higgs-boson mass, $\MH$~\cite{twoloopres,gamb}. 

In order to improve this situation, an exact evaluation of
electroweak two-loop contributions would be desirable, where no expansion 
in $\Mt$ or $\MH$ is made. In this paper this is illustrated for the 
example of two specific diagrams, for which the exact result is
compared with the result of an expansion in $\Mt$ up to next-to-leading
order.
Then exact results~\cite{sbaugw2}
recently obtained for the Higgs-mass dependence of $\Delta r$, i.e.\ the
relation between the gauge-boson masses, are briefly reviewed.

The MSSM provides the most predictive framework beyond the SM. While the
direct search for supersymmetric particles has not been successful yet,
the precision tests of the theory provide the possibility for
constraining the parameter space of the model and could eventually
allow to distinguish between the SM and Supersymmetry via their respective
virtual effects. In contrast to the SM case the predictions for $\De r$
and the Z-boson observables in the MSSM are known at one-loop order
only~\cite{susyprec}. In order to treat the MSSM at the same level of 
accuracy as the SM, higher-order contributions should be incorporated.
Recently the QCD corrections to the $\rho$~parameter in the MSSM
have been evaluated~\cite{susydelrhos,susydelrhol}. 
In this paper the exact result
for the gluonic contribution to $\De r$ is presented. It is compared
with the approximation based on the contribution to the
$\rho$~parameter.

\section{Comparison of exact calculation and expansion in the top-quark
mass}

The calculation of top-quark or Higgs-boson contributions to $\Delta r$
and other processes with light external fermions at low energies
requires in particular the evaluation of two-loop self-energies on-shell, 
i.e.\ at non-zero external momentum, while vertex and box contributions 
can mostly be reduced to vacuum integrals. The problems encountered in
such a calculation are due to the large number of contributing Feynman
diagrams, their complicated tensor structure, the fact that scalar
two-loop integrals are in general not expressible in terms of
polylogarithmic functions~\cite{ScharfDipl}, and due to the need for a
two-loop renormalization, which has not yet been worked out in full
detail.

\begin{figure}[ht]
\begin{center}
\mbox{
\psfig{figure=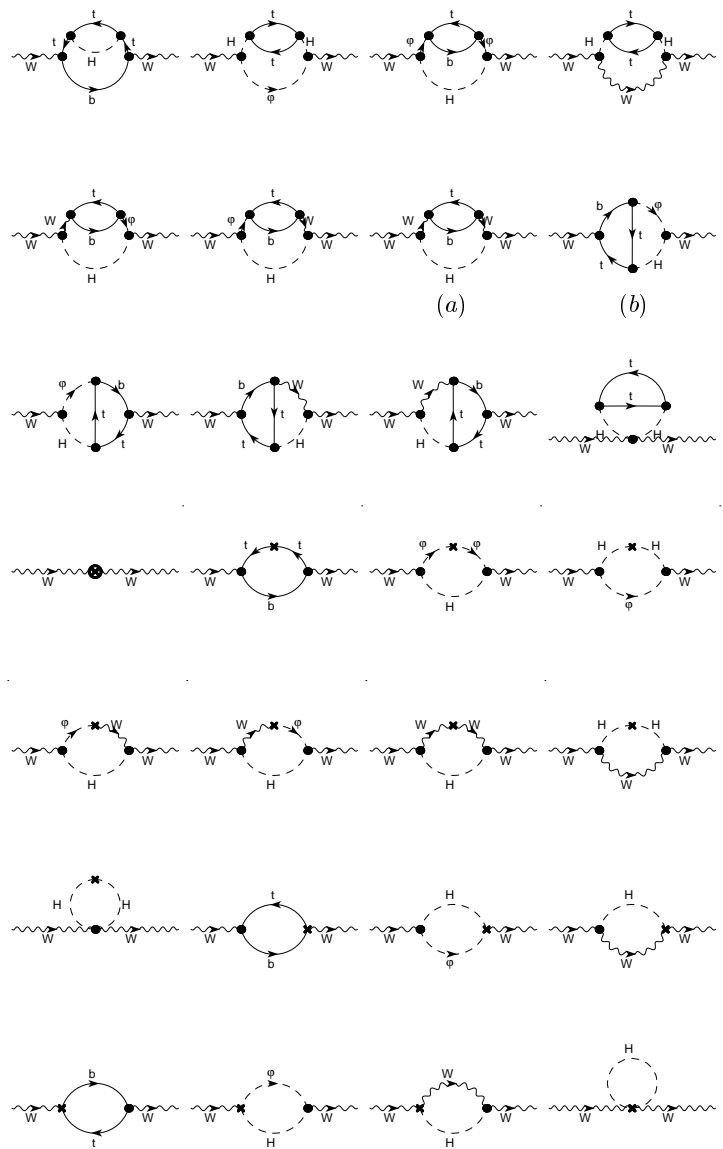,%width=9.5cm,height=6.5cm,
              height=15.5cm,
              bbllx=142pt,bblly=145pt,bburx=480pt,bbury=665pt}}
\caption{The two-loop top-quark corrections (including counterterm
graphs)
to the W-boson self-energy that depend on the Higgs-boson mass.
The diagrams $(a)$ and $(b)$ are marked for later reference.}
\label{fig:Wse}
\end{center}
\end{figure}

The methods used for the calculations discussed in this
paper have been outlined in \citere{sbaugw1}. The generation of the
diagrams and counterterm contributions is done with the help of the
computer-algebra program \fea\ \cite{fea}. As an example, in 
\reffi{fig:Wse} those diagrams contributing to the two-loop W-boson 
self-energy are given that contain both the top quark and the Higgs boson.
The corresponding counterterm graphs are also listed.

Making use of two-loop tensor-integral
decompositions, the generated amplitudes are reduced to a minimal set
of standard scalar integrals with the program \two~\cite{two}. The
renormalization is performed within the complete on-shell
scheme~\cite{onshell}, i.e.\
physical parameters are used throughout. The two-loop scalar integrals
are evaluated numerically with one-dimensional integral
representations~\cite{intnum}. These allow a very fast calculation of the
integrals with high precision without any approximation in the masses.

As an example, we consider the two diagrams marked as $(a)$ and $(b)$
in \reffi{fig:Wse} and compare the exact result with the result of an
expansion in the top-quark mass up to next to leading order, 
which takes into account terms of order $\Mt^4$ and $\Mt^2$ (see
\citere{gamb}). These diagrams represent the two main topologies for
the \se\ of the W~boson. 
The comparison of the two methods of
evaluation for just two diagrams can of course only provide an estimate 
of the relative difference that one can expect in the results for
physical observables. 
It has nevertheless the advantage that the direct comparison of
unrenormalized two-loop diagrams is not affected by differences in 
the renormalization schemes and in the treatment of subleading
higher-order corrections, which usually are present when 
predictions for physical observables are compared to each other.

The method for the asymptotic expansion in $\Mt$ has been described
in \citere{gamb}.%
\footnote{P.~Gambino has kindly provided us with his results for the
expansions of the two diagrams.}
It is performed in two regions, namely in the light Higgs region
(``light Higgs expansion''), for which 
$\MW, \MZ, \MH \ll \Mt$ is assumed, and in the heavy Higgs region 
(``heavy Higgs expansion''), which means $\MW, \MZ \ll \MH, \Mt$. 

We have checked that for asymptotically large
values of the top-quark mass the difference between the exact result
and the relevant expansion grows slower than $\Mt^2$, as it is required
as a matter of consistency. 

\begin{figure}[ht]
\begin{center}
\mbox{
\psfig{figure=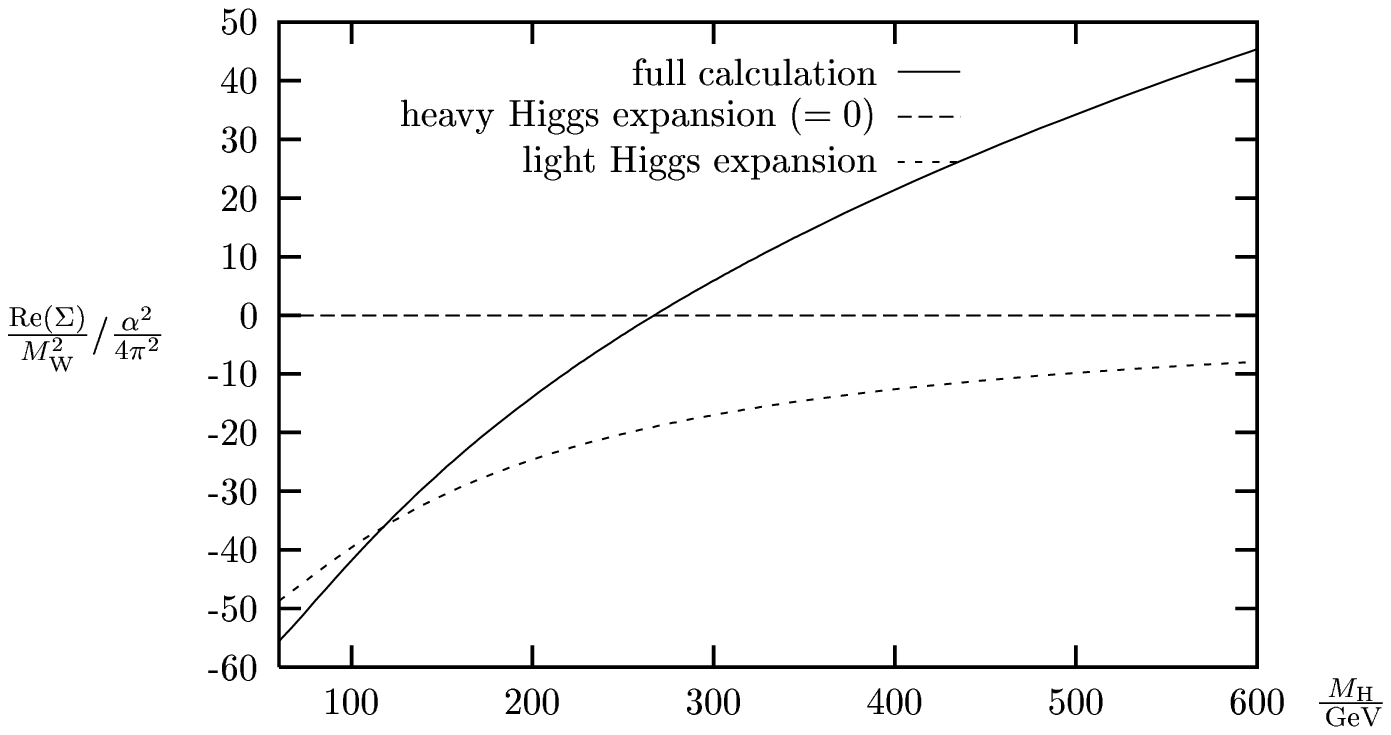,%width=9.5cm,height=6.5cm,
              width=10.5cm,
              bbllx=142pt,bblly=445pt,bburx=480pt,bbury=633pt}} \\[0.8cm]
\mbox{
\psfig{figure=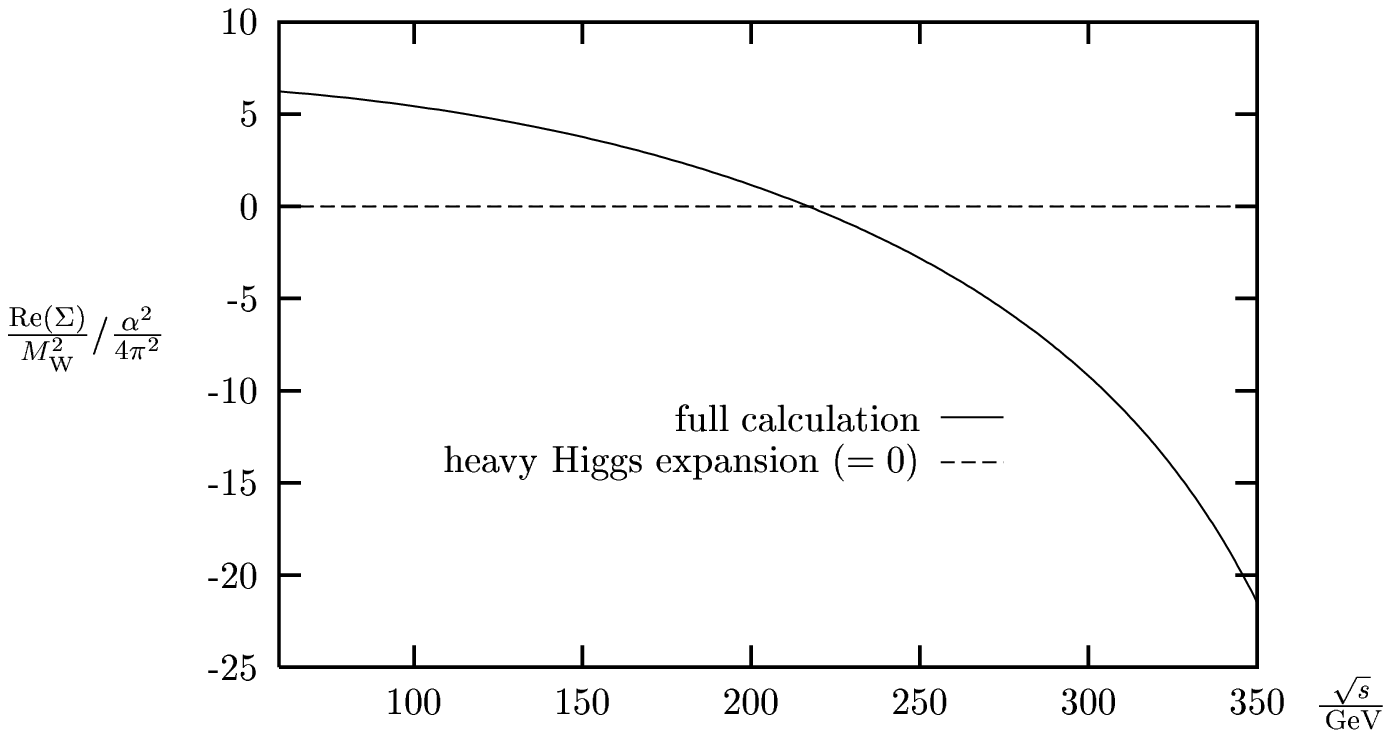,%width=9.5cm,height=6.5cm,
              width=10.5cm,
              bbllx=142pt,bblly=445pt,bburx=480pt,bbury=633pt}} \\[0.8cm]
\caption[]{Comparison between the exact result and the expansions up to
next-to-leading order in $\Mt$ for diagram $(a)$; $\Mt = 175$~GeV.
The upper plot shows the full result as well as the results for
the light Higgs and
heavy Higgs expansion (the latter is zero) as a function of $\MH$ for
$s = \MW^2$. The lower plot shows the full result and the heavy Higgs
expansion as a function of $\sqrt{s}$ for $\MH = 300$~GeV.
\label{fig:compa}
}
\end{center}
\end{figure}

\begin{figure}[ht]
\begin{center}
\mbox{
\psfig{figure=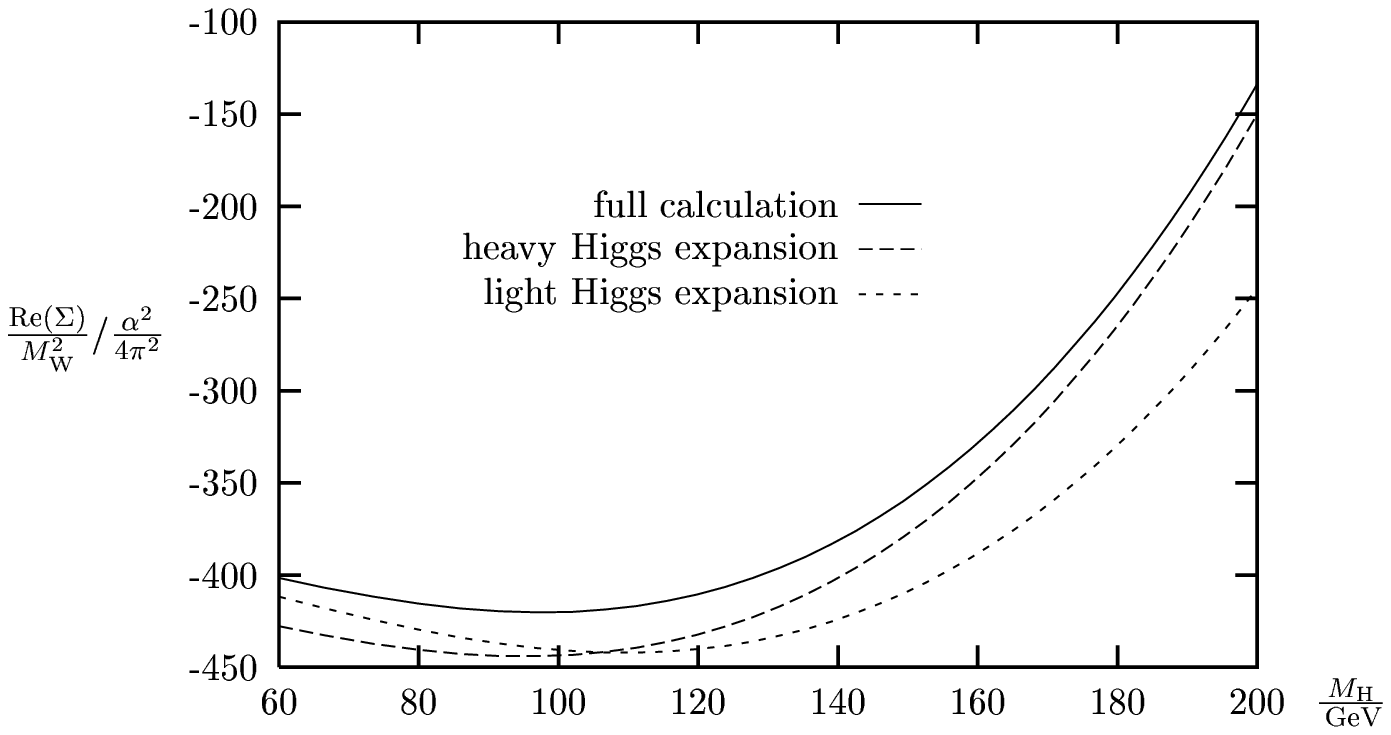,%width=9.5cm,height=6.5cm,
              width=10.5cm,
              bbllx=142pt,bblly=445pt,bburx=480pt,bbury=633pt}} \\[0.8cm]
\mbox{
\psfig{figure=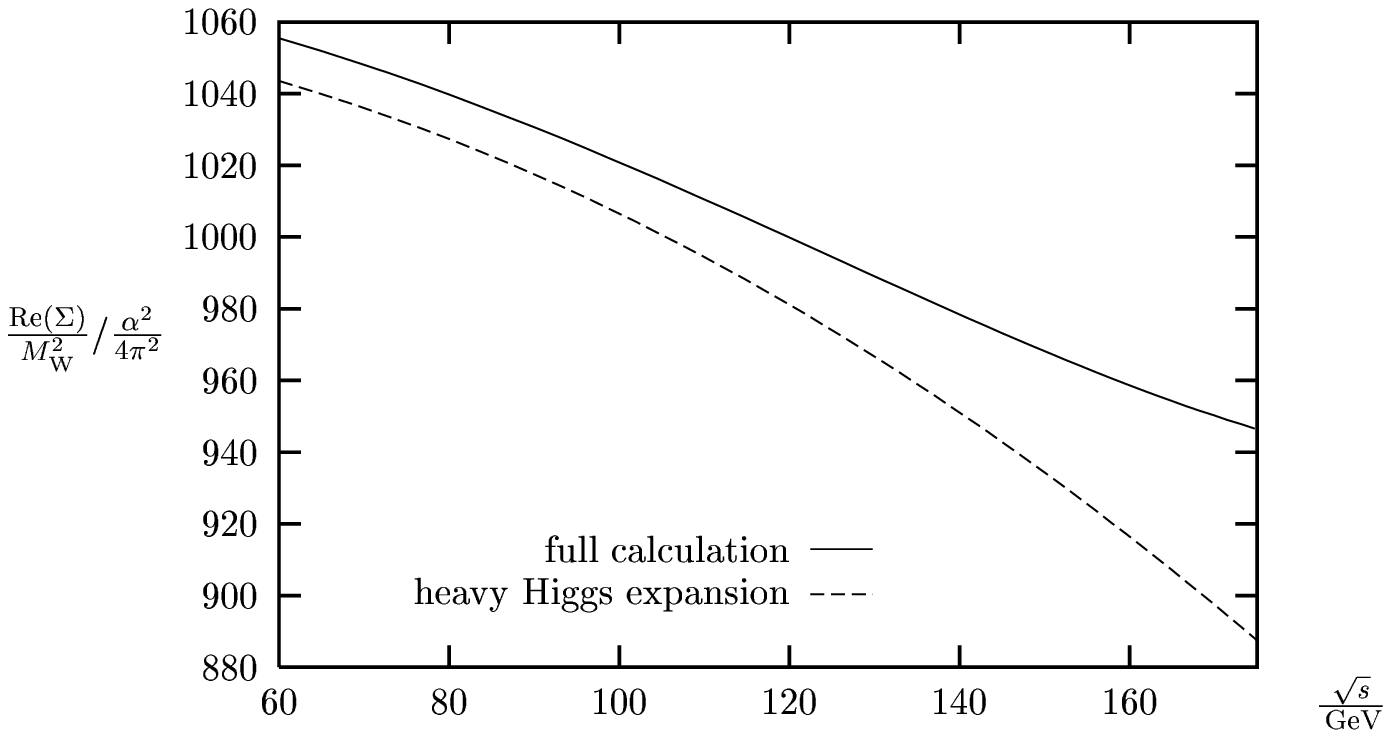,%width=9.5cm,height=6.5cm,
              width=10.5cm,
              bbllx=142pt,bblly=445pt,bburx=480pt,bbury=633pt}} \\[0.8cm]
\caption[]{Comparison between the exact result and the expansions up to
next-to-leading order in $\Mt$ for diagram $(b)$; $\Mt = 175$~GeV.
The upper plot shows the full result and the results for the
expansions as a function
of $\MH$ for $s = \MW^2$, while the lower plot shows the full result
and the heavy Higgs expansion as a function of $\sqrt{s}$ for $\MH =
300$~GeV.
\label{fig:compb}
}
\end{center}
\end{figure}

In the physical parameter region, i.e.\ for $\Mt = 175$~GeV, 
the comparison between the exact result and the expansions is displayed
in \reffi{fig:compa} and \reffi{fig:compb} for diagram $(a)$ and $(b)$,
respectively. The plots show the finite parts of the diagrams (divided
by $\MW^2$ and in units of $\frac{\alpha^2}{4 \pi^2}$) in
dimensional regularization and are given as a function of the 
Higgs-boson mass and of the external momentum $\sqrt{s}$.
In the latter plots $\MH = 300$~GeV has been used and only the curve
for the heavy Higgs expansion is shown since the light Higgs expansion 
is not valid for this value of $\MH$. 

For diagram $(a)$ the expansion in the heavy Higgs region yields zero,
and the deviations between the exact result and the expansions are
relatively large for most values of $\MH$ and $\sqrt{s}$. For diagram 
$(b)$ we find relatively good agreement between the full result and the
expansions. It should be noted, however, that the numerical
contribution of diagram $(b)$ is rather large so that in the final result
considerable cancellations can be expected.
The absolute deviation is for both diagrams of the order of
$10 \times \frac{\alpha^2}{4 \pi^2}$ to $50 \times \frac{\alpha^2}{4
\pi^2}$. 

%{\samepage

\section{Higgs-mass dependence of two-loop corrections to $\De r$}

The relation between the vector-boson masses in terms of the Fermi
constant $\GF$ reads~\cite{sirlin}
%}
\beq
\MW^2 \left(1 - \frac{\MW^2}{\MZ^2}\right) =
\frac{\pi \al}{\sqrt{2} \GF} \left(1 + \De r\right),
\eeq
where the radiative corrections are contained in the quantity $\De r$.
In the context of this paper we treat $\De r$ without resummations,
i.e.\ as being fully expanded up to two-loop order,
\beq
\De r = \De r_{(1)} + \De r_{(2)} + %{\cal O}(\al^3) .
\ldots \; .
\eeq
The theoretical predictions for $\De r$ are obtained by calculating
radiative corrections to muon decay.

We study the variation of the two-loop contributions to $\De r$ 
with the Higgs-boson mass by considering the subtracted quantity
\beq
\label{eq:DeltaRsubtr}
\De r_{(2), {\mathrm subtr}}(\MH) =
\De r_{(2)}(\MH) - \De r_{(2)}(\MH = 65\GeV),
\eeq
where $\De r_{(2)}(\MH)$ denotes the two-loop contribution to
$\De r$. Potentially large $\MH$-dependent contributions are the
corrections associated with the top quark, since the Yukawa coupling of
the Higgs to the top quark is proportional to $\Mt$, and the
contributions which are proportional to $\De\al$. We present here
exact results for the fermionic contributions, while the purely bosonic
corrections, which contain no specific source of enhancement, have been
estimated to give rise to only a relatively small shift in the 
W-boson mass of the order of $2$~MeV over the range
$65\, \GeV \leq \MH \leq 1\, \TeV$~\cite{sbaugw2}.

We begin with the Higgs-mass dependence of the two-loop top-quark
contributions and consider the quantity $\De r^{\mathrm top}_{(2),
{\mathrm subtr}}(\MH)$ which denotes the contribution of the top/bottom
doublet to $\De r_{(2), {\mathrm subtr}}(\MH)$. From the one-particle
irreducible diagrams obviously those graphs contribute to $\De
r^{\mathrm top}_{(2), {\mathrm subtr}}$ that contain both the top
and/or bottom quark and the Higgs boson. For the W~\se\ the relevant
graphs are the ones shown in \reffi{fig:Wse}. It is easy to see that
only two-point functions enter in $\De r^{\mathrm top}_{(2), {\mathrm
subtr}}(\MH)$, since all graphs where the Higgs boson couples to the
muon or the electron may safely be neglected. Although no two-loop
three-point function enters, there is nevertheless a contribution from
the two-loop and one-loop vertex counterterms. If the field
renormalization constants of the W~boson are included (which cancel in
the complete result), the vertex counterterms are separately finite.
The technically most complicated contributions arise from the mass and
mixing-angle renormalization. Since it is performed in the on-shell
scheme, the evaluation of the W- and Z-boson self-energies are required
at non-zero momentum transfer. 

The contribution of the terms proportional to $\De \al$ has the simple
form $\De r^{\De\al}_{(2), {\mathrm subtr}}(\MH) = 2 \De\al \, \De
r_{(1), {\mathrm subtr}}(\MH)$ and can easily be obtained by a proper
resummation of one-loop terms. The remaining fermionic contribution,
$\De r^{\mathrm lf}_{(2), {\mathrm subtr}}$,
is the one of the light fermions,
i.e.\ of the leptons and of the quark doublets of the first and second
generation,
which is not contained in $\De\al$. Its structure is analogous to
$\De r^{\mathrm top}_{(2), {\mathrm subtr}}$, 
but because of the negligible coupling of
the light fermions to the Higgs boson much less diagrams contribute.

\begin{figure}[ht]
\begin{center}
\mbox{}\\[0.2cm]
\mbox{
\psfig{figure=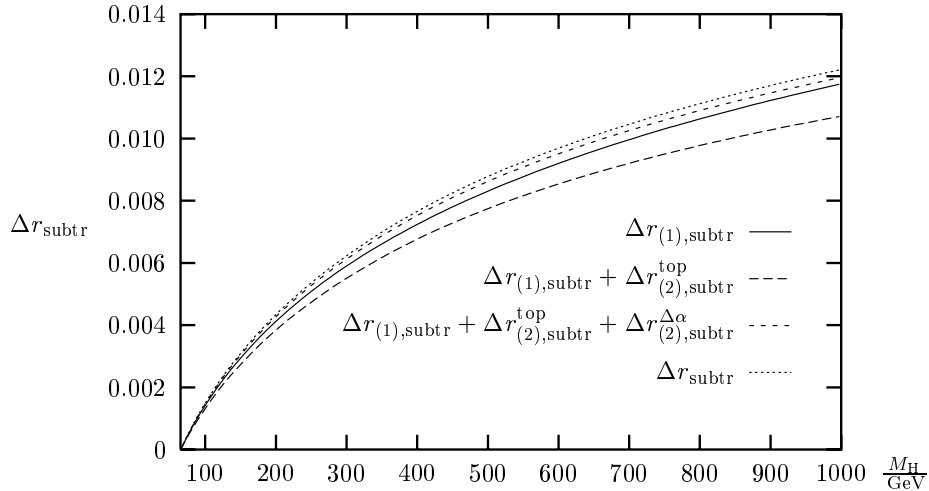,%width=9.5cm,height=6.5cm,
              width=10.5cm,
              bbllx=142pt,bblly=445pt,bburx=480pt,bbury=633pt}} \\[0.8cm]
\caption{One-loop and two-loop contributions to $\Delta r$
subtracted at $\MH=65\,\GeV$.
$\De r_{\mathrm subtr}$ is the result for the full one-loop and
fermionic
two-loop contributions to $\De r$, as defined in the text.
\label{fig:delr2}
}
\end{center}
\end{figure}

The total result for the one-loop and fermionic two-loop contributions
to $\De r$, subtracted at $\MH=65\,\GeV$, reads
\beq
\De r_{\mathrm subtr} \equiv \De r_{(1), {\mathrm subtr}} +
\De r^{\mathrm top}_{(2), {\mathrm subtr}} +
\De r^{\De\al}_{(2), {\mathrm subtr}} +
\De r^{\mathrm lf}_{(2), {\mathrm subtr}} .
\eeq
It is shown in \reffi{fig:delr2}, where separately also
the one-loop contribution
$\De r_{(1), {\mathrm subtr}}$, as well as
$\De r_{(1), {\mathrm subtr}} + \De r^{\mathrm top}_{(2), {\mathrm
subtr}}$, and
$\De r_{(1), {\mathrm subtr}} + \De r^{\mathrm top}_{(2), {\mathrm
subtr}} + \De r^{\De\al}_{(2), {\mathrm subtr}}$
are shown for $\Mt = 175.6 \GeV$.
The two-loop contributions $\De r^{\mathrm top}_{(2), {\mathrm
subtr}}(\MH)$ and $\De r^{\De\al}_{(2), {\mathrm subtr}}(\MH)$ turn out
to be of similar size and to cancel each other to a large extent.
In total, the inclusion of the higher-order contributions discussed
here
leads to a slight increase in the sensitivity to the Higgs-boson mass
compared to the pure one-loop result.

We have compared the result for
$\De r_{\mathrm subtr}^{{\mathrm top}, \De\al} \equiv
\De r_{(1), {\mathrm subtr}} + \De r^{\mathrm top}_{(2), {\mathrm
subtr}} + \De r^{\De\al}_{(2), {\mathrm subtr}}$
with the result obtained via an expansion in $\Mt$ up
to next-to-leading order, i.e.\
${\cal O}(\GF^2 \Mt^2 \MZ^2)$~\cite{gamb}. The results agree within
about $30 \%$ of $\De r^{\mathrm top}_{(2), {\mathrm subtr}}(\MH)$,
which amounts to a difference in $\MW$ of up to about 
$4$~MeV~\cite{sbaugw2}. 

In \refta{tab:DeltaMW} the shift in $\MW$ corresponding to $\De r_{\mathrm
subtr}(\MH)$, i.e.~the change in the theoretical prediction for $\MW$
when varying the Higgs-boson mass from $65\, \GeV$ to $1\, \TeV$, is
shown for three values of the top-quark mass, $\Mt = 170, 175, 180\,
\GeV$. The dependence on the precise value of $\Mt$ is rather mild,
which is expected from the fact that $\Mt$ enters here only at the
two-loop level and that $\De r^{\mathrm top}_{(2), {\mathrm
subtr}}(\MH)$ has a local maximum in the region of $\Mt = 175$~GeV
(see \citere{sbaugw2}).

\btab
$$
\barr{|c||c|c|c|c|c|c|c|c|} \hline
\MH/\GeV & 65 & 100 & 200 & 300 & 400 & 500 & 600 & 1000 \\ \hline
\hline
\Mt=170\,\GeV &
0 & -22.6 & -65.8 & -94.5 & -116 & -133 & -147 & -185  \\ \hline
\Mt=175\,\GeV &
0 & -22.8 & -66.3 & -95.2 & -117 & -134 & -148 & -187  \\ \hline
\Mt=180\,\GeV &
0 & -23.0 & -66.8 & -96.0 & -118 & -135 & -149 & -188  \\ \hline
\earr
$$
\caption{The shift in MeV in the %dependence of the %shift in $\MW$ on the
theoretical prediction for $\MW$ caused by varying the Higgs-boson mass
in the interval $65\, \GeV \leq \MH \leq 1\, \TeV$
for three values of $\Mt$.
\label{tab:DeltaMW}}
\etab

\section{Gluonic corrections to $\De r$ in the MSSM}

In the MSSM, 
the leading contributions of scalar quarks to $\De r$ and the leptonic
Z-boson observables enter via the $\rho$~parameter.
The contribution of squark
loops to the $\rho$~parameter can be written in terms of the transverse
parts of the W- and Z-boson \ses\ at zero momentum-transfer,
\beq
\Delta \rho =
\frac{\Sigma^{\PZ}(0)}{\MZ^2} - \frac{\Sigma^{\PW}(0)}{\MW^2} .
\eeq
The one-loop result for the stop/sbottom doublet in the MSSM
reads~\cite{R6rho}
\beqar
\Delta \rho ^{\rm SUSY}_0 &=& \frac{3 \GF}{8 \sqrt{2} \pi^2} \left[ -
\sin^2 \theta_{\tilde{\Pt}} \cos^2 \theta_{\tilde{\Pt}}
F_0\left( m_{\tilde{\Pt}_1}^2,  m_{\tilde{\Pt}_2}^2 \right)
\right. \nonumber \\
&& \left. + \cos^2 \theta_{\tilde{\Pt}} F_0\left( m_{\tilde{\Pt}_1}^2,
m_{\tilde{\Pb}_L}^2 \right) + \sin^2 \theta_{\tilde{\Pt}} F_0\left(
m_{\tilde{\Pt}_2}^2,  m_{\tilde{\Pb}_L}^2 \right) \right] ,
\label{eq:lett5}
\eeqar
where $\theta_{\tilde{\Pt}}$ is the mixing angle in the stop sector.
Only the left-handed sbottom mass, $m_{\tilde{\Pb}_L}$, contributes
since mixing in the sbottom sector has been neglected.
The function $F_0(x,y)$ has the form
\beq
F_0(x,y)= x+y - \frac{2xy} {x-y} \log \frac{x}{y}  .
\label{eq:lett3}
\eeq
It vanishes if the squarks are degenerate in mass, $F_0(m^2, m^2)=0$. 
In the limit of a large mass splitting between the 
squarks it is proportional to the heavy squark mass squared,
$F_0(m^2,0)=m^2$. This is in analogy to the case of the top/bottom
doublet in the SM~\cite{velt},
\beq
\Delta \rho_0^{\rm SM} = \frac{3 \GF}{8 \sqrt{2} \pi^2}
F_0(\Mt^2,\Mb^2) \approx
\frac{3 \GF \Mt^2}{8 \sqrt{2} \pi^2} .
\eeq

Since the contribution of a squark doublet vanishes
if all masses are degenerate, in most SUSY scenarios only the third 
generation contributes. While the scalar partners of the light quarks
are almost mass degenerate in most SUSY scenarios, in the third
generation the top-quark mass enters the mass matrix of the scalar
partners of the top quark (see \refeq{eq:mm1} below). 
It can give rise to a large mixing in the
stop sector and to a large splitting between the stop and sbottom
masses.

Recently the QCD corrections to the $\rho$ parameter in the MSSM have
been evaluated~\cite{susydelrhos,susydelrhol}. 
The two-loop Feynman diagrams of the
squark loop contributions to $\De\rho$ at ${\cal O}(\alpha \alpha_s)$
can be divided into diagrams in which a gluon is exchanged, into 
diagrams with gluino exchange, and into pure scalar diagrams. 
After the inclusion of the corresponding counterterms the
contribution of the pure scalar diagrams vanishes and the other two
sets are separately ultraviolet finite and gauge-invariant
(see \citere{susydelrhol}). 

The result for the gluon-exchange contribution is given by a simple 
expression resembling the one-loop result of \refeq{eq:lett5},
\beqar
\Delta \rho ^{\rm SUSY}_{1, {\rm gluon}} &=& 
\frac{\GF \alpha_s}{4 \sqrt{2} \pi^3} \left[
- \sin^2\theta_{\tilde{\Pt}} \cos^2\theta_{\tilde{\Pt}}
F_1\left( m_{\tilde{\Pt}_1}^2,  m_{\tilde{\Pt}_2}^2 \right) \right.
\nonumber \\
&&\left. + \cos^2 \theta_{\tilde{\Pt}} F_1 \left( m_{\tilde{\Pt}_1}^2,
m_{\tilde{\Pb}_L}^2 \right)
+\sin^2 \theta_{\tilde{\Pt}}  F_1 \left(
m_{\tilde{\Pt}_2}^2,  m_{\tilde{\Pb}_L}^2 \right) \right] .
\label{eq:delrhoglu}
\eeqar
The two-loop function $F_1(x,y)$ is given in terms of
dilogarithms by
\beqar
F_{1}(x,y) &=& x+y- 2\frac{xy}{x-y} \log \frac{x}{y} \left[2+
\frac{x}{y} \log \frac{x}{y} \right] \nonumber \\
&& {} +\frac{(x+y)x^2}{(x-y)^2}\log^2 \frac{x}{y}
-2(x-y) {\rm Li}_2 \left(1-\frac{x}{y} \right) .
\eeqar
It is symmetric in the interchange of $x$ and $y$.
It vanishes for degenerate masses, $F_1(m^2,m^2)=0$, while in the case of
large mass splitting it increases with the heavy scalar quark mass
squared: $F_1 (m^2,0) = m^2( 1 +\pi^2/3)$. 

It is remarkable that contrary to the Standard Model case~\cite{lb}, 
\beq
\Delta \rho ^{\rm SM}_1 = - \Delta \rho_0^{\rm SM} \,
\frac{2}{3} \frac{\alpha_s}{\pi} (1+ \frac{\pi^2}{3} ) ,
\eeq
where the QCD corrections are negative and screen the one-loop value, 
$\Delta \rho ^{\rm SUSY}_{1, {\rm gluon}}$ enters with the same sign as
the one-loop contribution. It therefore enhances the sensitivity in the
search of the virtual effects of scalar quarks in high-precision
electroweak measurements.

The analytical expression for the gluino-exchange contribution is much
more complicated than \refeq{eq:delrhoglu}. In general the gluino-exchange 
diagrams give smaller contributions compared to gluon exchange. 
Only for gluino and squark masses close to the experimental
lower bounds they compete with the gluon-exchange contributions. 
For higher values of $m_{\tilde{\mathrm g}}$, the contribution
decreases rapidly since the gluino decouples in the large-mass limit.

The leading contribution to $\De r$ in the MSSM can be approximated by
the contribution to the $\rho$~parameter according to
\beq
\De r \approx - \frac{\cw^2}{\sw^2} \De \rho,
\label{eq:delrhoappr}
\eeq
where $\cw^2 = 1 - \sw^2 = \MW^2/\MZ^2$.

We have derived the exact result for the gluon-exchange contribution to
$\De r$, so that the accuracy of the approximation
\refeq{eq:delrhoappr} can be tested in this case. The gluon-exchange
correction to the contribution of a squark doublet to $\De r$ is given
by
\beq
\Delta r^{\rm SUSY}_{\rm gluon} = \Pi^{\gamma}(0) -
\frac{\cw^2}{\sw^2} \left(\frac{\de \MZ^2}{\MZ^2} - 
\frac{\de \MW^2}{\MW^2} \right) +
\frac{\Sigma^\PW(0) - \de \MW^2}{\MW^2},
\label{eq:deltrglu}
\eeq
where $\de \MW^2 = \Re \Sigma^{\PW}(\MW^2)$,
$\de \MZ^2 = \Re \Sigma^{\PZ}(\MZ^2)$, and $\Pi^{\gamma}$, 
$\Sigma^{\PW}$, and $\Sigma^{\PZ}$ denote the transverse parts of the
two-loop gluon-exchange contributions to the photon vacuum polarization
and the W-boson and Z-boson \ses, respectively, 
which all are understood to contain the subloop renormalization.

In contrast to the $\rho$~parameter, the evaluation of
\refeq{eq:deltrglu} requires the calculation of two-loop two-point
functions at non-zero momentum transfer. Since the gluon is massless an
analytical result in terms of polylogarithmic functions can be
obtained. For deriving this result we have used the expressions for the
relevant two-loop scalar integrals given in \citere{lbg}.

\begin{figure}[htb]
\begin{center}
\mbox{
\psfig{figure=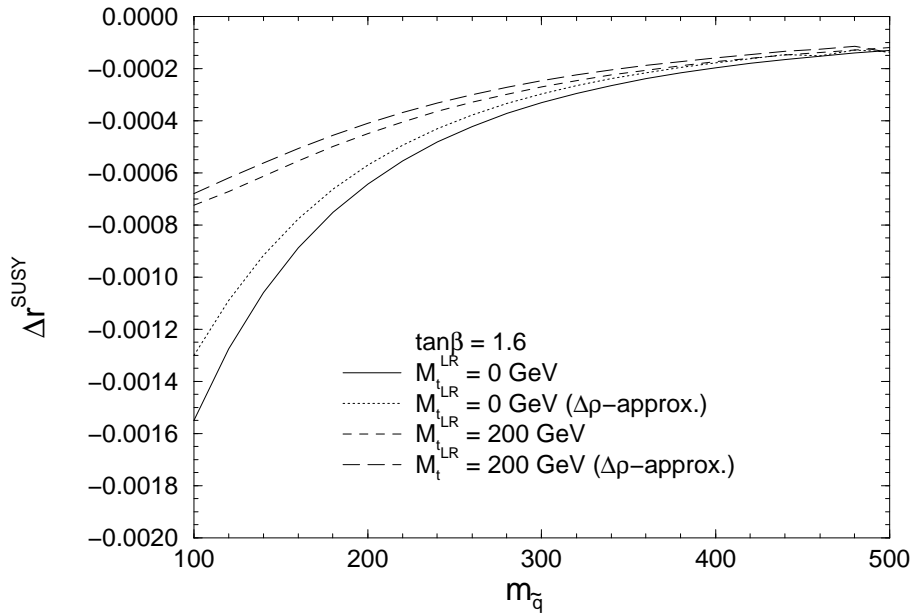,width=12cm,%height=6.5cm,
              bbllx=25pt,bblly=85pt,bburx=550pt,bbury=445pt}}
\parbox{12cm}{
\caption[]{Contribution of the gluon-exchange diagrams to
$\Delta r^{\rm SUSY}$ as a function of the common scalar mass parameter
$m_{\tilde q}$ for the scenario of no mixing 
($M^{LR}_t=0$) and of maximal mixing ($M_t^{LR}=200$~GeV)
in the stop sector. The exact result is compared with the approximation
derived from the contribution of $\De\rho$.
\label{fig:deltrglu}
}}
\end{center}
\end{figure}

The gluon-exchange correction to the contribution of the stop/sbottom loops
to $\De r$ is shown in \reffi{fig:deltrglu} together with the 
$\De\rho$ approximation according to \refeq{eq:delrhoappr}
as a function of the common
scalar mass parameter
\beq
m_{\tilde q} =
M_{\tilde{t}_{L}}=M_{\tilde{t}_{R}}=M_{\tilde{b}_{L}} =
M_{\tilde{b}_{R}} .
\label{eq:susyrel}
\eeq
The $M_{\tilde{q}_i}$ are the soft SUSY breaking parameters
appearing in the stop and sbottom mass matrices,
\beqar
{\cal M}^2_{\tilde{t}} &=& \left(
  \begin{array}{cc} M_{\tilde{t}_L}^2 + \Mt^2 + 
d_{\tilde{t}_L}
& \Mt \, M^{LR}_t
\\
\Mt \, M^{LR}_t & M_{\tilde{t}_R}^2 + \Mt^2 + 
d_{\tilde{t}_R}
  \end{array}
\right)
\label{eq:mm1} \\
{\cal M}^2_{\tilde{b}} &=& 
\left(
  \begin{array}{cc} M_{\tilde{b}_L}^2 + \Mb^2 + 
d_{\tilde{b}_L}
& \Mb \, M^{LR}_b
\\
\Mb \, M^{LR}_b & M_{\tilde{b}_R}^2 + \Mb^2 +
d_{\tilde{b}_R}
  \end{array}
\right) ,
\label{eq:mm2}
\eeqar
with $M^{LR}_t = A_t - \mu \, \CTb$, $M^{LR}_b = A_b - \mu \tb$,
and the $d_{\tilde{q}_i}$ are specified e.g.\ in \citere{susydelrhol}.
In this scenario, the
scalar top mixing angle is either very small, $\theta_{\tilde t} \sim
0$, or almost maximal, $ \theta_{\tilde t} \sim  -\pi/4$, in most of
the MSSM parameter space. The plots are shown for the two cases
$M^{LR}_t=0$ (no mixing) and $M_t^{LR}=200$~GeV (maximal mixing) for
$\tb=1.6$.

The two-loop contribution $\Delta r^{\rm SUSY}_{\rm gluon}$ is of the
order of 10--15\% of the one-loop result. It yields a shift in
the W-boson mass of up to $20$~MeV for low values of $m_{\tilde q}$ in
the no-mixing case. If the parameter $M^{LR}_t$
is made large or the relation \refeq{eq:susyrel} is relaxed, much
bigger effects are possible~\cite{susydelrhol}.
As can be seen in \reffi{fig:deltrglu}, the $\De\rho$ contribution 
approximates the full result rather well. The two results agree within 
10--15\%. 

{}From the good agreement between full result and $\De\rho$
approximation for the gluonic contributions one can expect for the
gluino-exchange correction, whose contribution to $\De\rho$ is in
general smaller than the gluonic part, that it should be sufficiently
well approximated by the $\De\rho$ contribution.

\section{Conclusions}

Recent results obtained at two-loop order for the relation between the 
gauge-boson masses in the Standard Model and the MSSM have been reviewed. 
For the example of two typical SM diagrams contributing to the W-boson
\se\ exact two-loop results have been evaluated and compared with an 
asymptotic expansion in the top-quark mass up to next-to-leading order. 
While for asymptotically large values of the top-quark mass the consistency
between the two results has been verified, for the physical value of the
top-quark mass the relative difference between exact result and
expansion can be sizable. 

In the Standard Model, the Higgs-mass dependence of the relation
between the gauge-boson masses has been studied. Exact two-loop results
have been given for the fermionic contributions, while the 
extra shift coming from
the purely bosonic two-loop corrections is expected to be negligible.  
As far as the Higgs-mass dependence of $\De r$ is concerned, with the
results presented here the theoretical uncertainty due to unknown
higher-order corrections should now be under control.

In Supersymmetry, the exact result for the gluon-exchange correction to
the contribution of squark loops to $\De r$
has been presented. It gives rise to a shift in the W-boson
mass of up to $20$~MeV. The result has been compared with the recently
obtained leading contribution entering via the $\rho$~parameter, and
good agreement has been found.

\section*{Acknowledgements}

The author thanks K.~Ko\l{}odziej and the other organizers of the
Ustron Conference for the invitation, the excellent organization and
their kind hospitality during the Conference. I am grateful to
my collaborators S.~Bauberger, A.~Djouadi, P.~Gambino,
S.~Heinemeyer, W.~Hollik and C.~J\"unger, with whom the results
presented here have been worked out.

\end{document}